\documentclass[
reprint,
amsmath,amssymb,
aps,pra
]{revtex4-2}

\usepackage{graphicx}
\usepackage{dcolumn}
\usepackage{bm}
\usepackage{color}
\usepackage{comment}

\usepackage{hyperref}
\hypersetup{
colorlinks=true,
citecolor=blue,
linkcolor=blue,
urlcolor=blue
}

\begin{document}

\preprint{APS/123-QED}

\title{
Anomalous minimization for critical velocity of superflow along a step potential
}

\author{Akihiro Kanjo$^{1}$}
\author{Hiromitsu Takeuchi$^{1,2}$}
\affiliation{
$^1$ Department of Physics, Osaka Metropolitan University, 3-3-138 Sugimoto, Osaka 558-8585, Japan \\
$^2$ Nambu Yoichiro Institute of Theoretical and Experimental Physics (NITEP), Osaka Metropolitan University, 3-3-138 Sugimoto, Osaka 558-8585, Japan
}

\date{\today}


\begin{abstract}
To reveal a microscopic mechanism for the anomalous minimization and dependence of the superfluid critical velocity on a moving obstacle potential in a atomic Bose-Einstein condensate [\href{https://link.aps.org/doi/10.1103/PhysRevA.91.053615}{Phys.~Rev.~A \textbf{91}, 053615 (2015)}],
we introduce a considerably simplified model of superflow along a step potential.
The energy spectrum and wave functions of the lowest-energy excitations in this system are well described by the semi-classical analysis based on the Bogoliubov theory.
We found that the critical velocity is minimized and becomes zero when the potential height equals the hydrostatic chemical potential, which corresponds to the critical point of the local condensation phase transition inside the step potential.
In a finite-size system, the critical velocity $v_\mathrm{c}$ obeys a power-law scaling  with the system size $L_x$ as $v_\mathrm{c}\propto L_x^{-0.963}$.
This criticality provides an explanation of the power-law scaling of the minimum critical velocity observed in the experiment.
\end{abstract}

\maketitle


\section{Introduction}\label{sec:introduction}
A superfluid flows without friction below a critical velocity~\cite{Pitaevskii2016a}.
According to the Landau criterion of superfluidity~\cite{Landau1941a}, the critical velocity is given by $v_\mathrm{c}=\min_p[\epsilon(p)/p]$, where $\epsilon(p)$ is an energy spectrum of an elementary excitation with momentum $p$.
In liquid helium II, the Landau spectrum determines the critical velocity, above which the excitations such as phonons and rotons are spontaneously emitted, leading to dissipative flow~\cite{Landau1941a,Feynman1956a}.
In addition, quantum vortices play a crucial role in the breakdown of superfluidity through nucleation of vortex rings~\cite{Feynman1955a}, phase slips induced by vortex motion~\cite{Anderson1966a}, and growth of remnant vortices pinned to channel boundaries~\cite{Donnelly1991a}.
In homogeneous Bose--Einstein condensates (BECs) of dilute atomic gases, the Bogoliubov spectrum yields $v_\mathrm{c}=c_\mathrm{s}$, where $c_\mathrm{s}$ is the speed of sound ~\cite{Bogoliubov1947a}.
In practice, however, the critical velocity is highly sensitive to the spatial inhomogeneity of the condensates and to the geometry of obstacles.
Indeed, previous studies using a moving obstacle potential have experimentally~\cite{Raman1999a,Onofrio2000a,Inouye2001a,Neely2010a,Desbuquois2012a,Kwon2014a,Kwon2016a} and theoretically~\cite{Frisch1992a,Huepe2000a,Rica2001a,Aftalion2003a,Pham2005a,Pinsker2014a,Kunimi2015a,Kiehn2022a,Kwak2023a,Huynh2024a,Christenhusz2025a} demonstrated the lower critical velocities, at which vortex nucleation marks the onset of energy dissipation.
Consequently, a quantitative prediction of the critical velocity remains a challenging task.

In the experiment by Kwon \textit{et al.}~\cite{Kwon2015a}, the critical velocity was measured as a function of the peak height $V_0$ of a repulsive Gaussian laser beam in a highly oblate BEC.
Remarkably, they observed that the critical velocity $v_\mathrm{c}$ exhibits a sharp minimum at a certain critical height close to the chemical potential $\mu$, independent of the beam width.
For $V_0<\mu$, $v_\mathrm{c}$ corresponds to the local speed of sound, which decreases together with the local density at the center of the obstacle~\cite{Kwon2015a,Kwak2023a}.
Similar reductions of $v_\mathrm{c}$ have been demonstrated in one-dimensional systems with delta-function~\cite{Hakim1997a}, Gaussian~\cite{Albert2008a,Huynh2022a}, rectangular~\cite{Takahashi2009a,Watanabe2009a,Paris2017a,Huynh2023a}, and periodic potentials~\cite{Watanabe2009a,Frisch2024a}.
For $V_0\gg\mu$ with a steep potential slope, $v_\mathrm{c}$ converges to a constant value, consistent with $v_\mathrm{c}\approx0.37c_\mathrm{s}$ for a hard cylinder~\cite{Huepe2000a,Rica2001a,Pham2005a}.
From a hydrodynamic perspective, the potential flow theory~\cite{Batchelor1967a} predicts that $v_\mathrm{c}$ is determined by the local speed of sound in the vicinity of a hard obstacle such as disks~\cite{Rica2001a,Pham2005a,Huynh2024a}, ellipses~\cite{Stagg2014a,Stagg2015a}, thin plates~\cite{Kokubo2024a,Kokubo2025a}, and airfoils~\cite{Musser2019a}.
Despite these experimental and theoretical studies, 
the anomalous minimization of $v_\mathrm{c}$ has not been explained microscopically in the context of the Landau criterion based on Bogoliubov analysis.

Recent advances in experimental techniques provide an ideal platform to address this problem.
First, uniform superfluids can be realized in cold atomic gases trapped in a cylindrical optical box~\cite{Gaunt2013a,Navon2021a,Kwon2021a,Hernandez2024a}.
Second, digital micromirror devices (DMDs), which allow spatial modulation of the amplitude of a laser beam, enable flexible control of the shape and height of external potentials~\cite{Gauthier2016a,Navon2021a}.
These developments will make it possible to elucidate the intrinsic effects of obstacle geometry and height on the critical velocity.

In this study, we reveal the microscopic mechanism by which the critical velocity $v_\mathrm{c}$ is minimized
through the use of a simple model of a BEC flowing along a step potential [Fig.~\ref{fig:FIG1}(a)].
While in Ref.~\cite{Kwon2015a} there are superflow both along and toward (and also through) the two-dimensional Gaussian obstacle, we here consider only a superflow along the step potential.
By investigating the stationary superflow, we find that a certain potential height corresponds the critical point of the local condensation phase transition inside the step potential.
Before evaluating $v_\mathrm{c}$ based on the Landau criterion, we numerically investigate the lowest-energy excitations.
The semi-classical theory successfully describes characteristic wave functions of the excitations and provides the theoretical dispersion relations of the excitation energy. 
Based on this analysis, we demonstrate that $v_\mathrm{c}$ is minimized and becomes even zero in an infinite system, consistent with the experiment by Kwon \textit{et al.}~\cite{Kwon2015a}.
This minimization is related to the criticality of the local condensation phase transition.

This paper is organized as follows.
In Sec.~\ref{sec:stationary superflow}, we introduce the basic formulation and investigate stationary superflow along a step potential.
In Sec.~\ref{sec:lowest-energy excitations}, we show typical dispersion relations and wave functions of the lowest-energy excitations based on the Bogoliubov theory.
In Sec.~\ref{sec:semi-classical analysis}, we formulate the semi-classical theory for bosonic quasiparticles.
The main results for the critical velocity are presented in Sec.~\ref{sec:critical velocity}.
Section~\ref{sec:summary} is devoted to a summary and discussion.

\section{Stationary superflow along a step potential}\label{sec:stationary superflow}

\subsection{Basic formulation}\label{sec:basic formulation}
\begin{figure}[t]
\centering
\includegraphics[width=1.0\columnwidth]{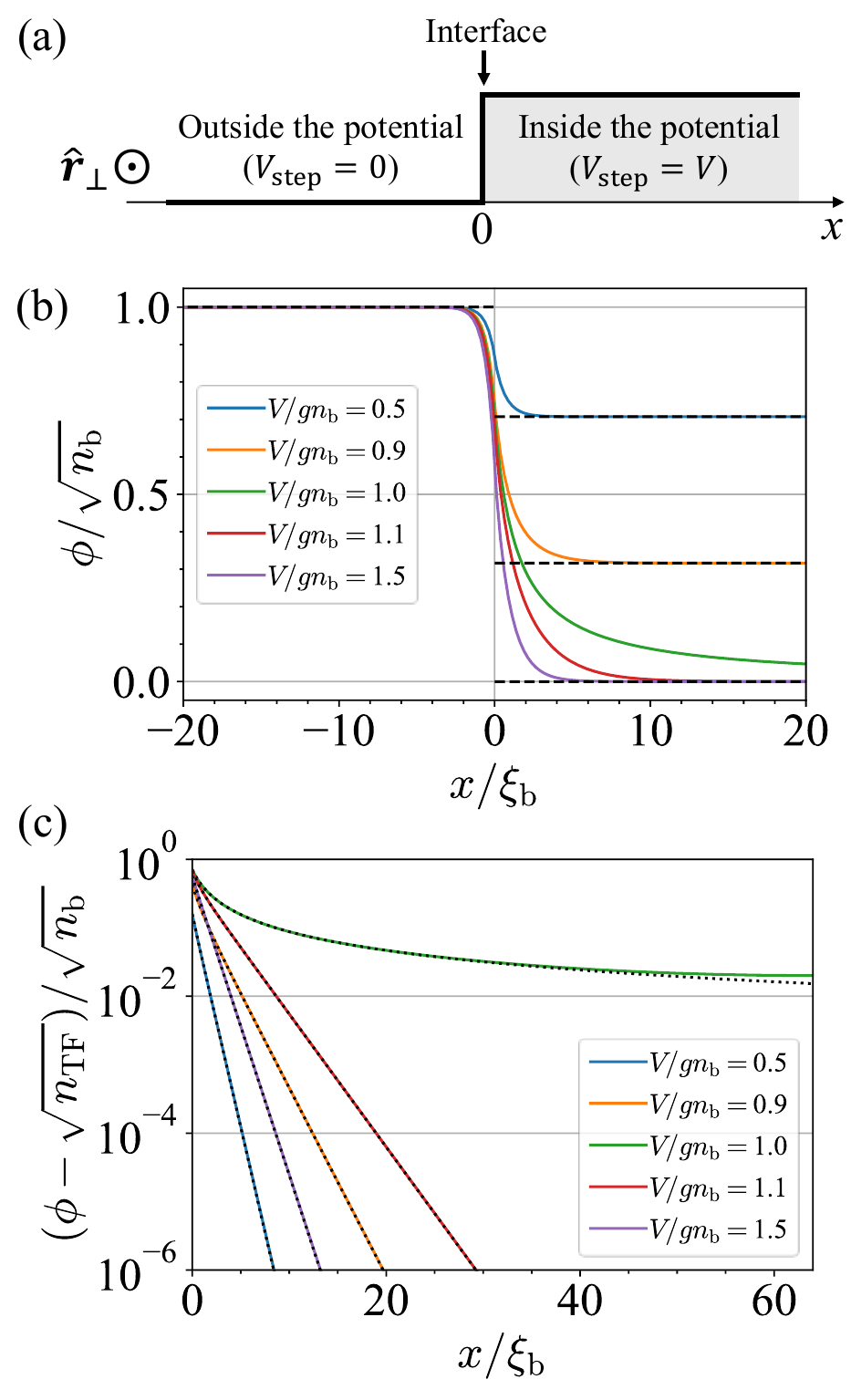}
\caption{
(a) Schematic of our system with a step potential $V_\mathrm{step}(x)$ [Eq.~\eqref{eq:step potential}].
(b) Profiles of the stationary states $\phi$ of a condensate flowing along the step potential 
for $V/gn_\mathrm{b}=0.5,0.9,1,1.1,$ and $1.5$.
Black dashed lines indicate $\sqrt{n_\mathrm{TF}}$, where $n_\mathrm{TF}$ is the TF density profile [Eq.~\eqref{eq:TF desity profile}].
(c) Corresponding profiles of $\phi-\sqrt{n_\mathrm{TF}}$ shown with a logarithmic vertical axis.
Black dotted curves indicate the analytical solutions of Eq.~\eqref{eq:time-independent GPE} in an infinite system shown in Fig.5 of Ref.~\cite{Takahashi2009a}.
}
\label{fig:FIG1}
\end{figure}

We consider a BEC described by the complex order parameter $\Psi(\bm{r},t)$ at zero temperature.
In the Gross-Pitaevskii (GP) model~\cite{Gross1961a,Pitaevskii1961a}, the mean-field Lagrangian is given by $\mathcal{L}=\int d^3x \, i\hbar\Psi^*\partial_t\Psi-\mathcal{F}$ with the energy functional
\begin{equation}
    \mathcal{F}
    =\int d^3x \, 
     \left\{
      \frac{\hbar^2}{2m}|\nabla\Psi|^2
      +(V_\mathrm{step}-\mu)|\Psi|^2
      +\frac{g}{2}|\Psi|^4
     \right\}.
    \label{eq:energy functional}
\end{equation}
Here, the atomic mass $m$, the chemical potential $\mu$, the interaction constant $g$, and the reduced Planck constant $\hbar$ are used.
The external potential $V_\mathrm{step}$ represents a step potential defined as 
\begin{equation}
    V_\mathrm{step}(x)=
    \begin{cases}
        V\geq0 & (\text{for} \ x\geq0) \\
        0      & (\text{for} \ x<0   )
    \end{cases}.
    \label{eq:step potential}
\end{equation}
As illustrated in Fig.~\ref{fig:FIG1}(a), we refer the regions $x>0$, $x=0$, and $x<0$ as ``inside the potential'', ``interface'', and ``outside the potential'', respectively

We consider a stationary superflow along the step potential at a constant velocity $\bm{v}=-v\hat{\bm{r}}_\perp$, where $\hat{\bm{r}}_\perp$ denotes the unit vector normal to the $x$-axis.
The stationary state $\Psi=\phi(x)e^{im\bm{v}\cdot\bm{r}/\hbar}$ satisfies the time-independent GP equation
\begin{equation}
    \left(
     -\frac{\hbar^2}{2m}\frac{d^2}{dx^2}
     +V_\mathrm{step}-\mu_v+g\phi^2
    \right)\phi=0.
    \label{eq:time-independent GPE}
\end{equation}
Here, $\mu_v\equiv\mu-mv^2/2$ is the hydrostatic chemical potential, named after the hydrostatic pressure in quantum hydrodynamics~\cite{Tsubota2013a}.
Our system is Galilean invariant and thus the change of Eq.~\eqref{eq:time-independent GPE} under the Galilean transformation reduces to the shift of the chemical potential $\mu\to\mu_v$.
Because $\mu_v$ is fixed and independent on $\bm{v}$ through the boundary condition $\mu_v=gn_\mathrm{b}$ at the bulk with the bulk density $n_\mathrm{b}$, we replace $\mu_v$ with $gn_\mathrm{b}$ in the following discussions
By an appropriate choice of the global phase, $\phi$ can be taken to be real.
According to Eq.~\eqref{eq:energy functional}, the healing length outside the potential is given by $\xi_\mathrm{b}=\hbar/\sqrt{mgn_\mathrm{b}}$, while inside the potential it is given by $\xi(V)=|1-V/ gn_\mathrm{b}|^{-1/2}\xi_\mathrm{b}$.
These quantities coincide with the correlation lengths in the mean field approximation~\cite{Stanley1987a,Landau2013b}.
In the following, we take $\xi_\mathrm{b}$, $\hbar/ gn_\mathrm{b}$, and $\sqrt{n_\mathrm{b}}$ as the unit of length, time, and wave function, respectively.

\subsection{Profiles of the stationary states}\label{sec:profiles of the stationary states}
We investigate the $V$-dependence of the stationary states $\phi$.
Figure~\ref{fig:FIG1}(b) shows the typical profiles of $\phi$, obtained numerically by minimizing Eq.~\eqref{eq:energy functional} under the Neumann boundary conditions at $x=\pm L_x/2$. 
The system size $ L_x$ is chosen to be sufficiently large, with $L_x=128\xi_\mathrm{b}$.
See Appendix~\ref{app:numerical methods} for details on the numerical method. 
For comparison, we also plot the square of a Thomas-Fermi (TF) density profile~\cite{Pethick2008a} 
\begin{equation}
    n_\mathrm{TF}(x)=
    \begin{cases}
        (1-V_\mathrm{step}/ gn_\mathrm{b})n_\mathrm{b} & (\text{for} \ V_\mathrm{step}\leq gn_\mathrm{b}) \\
        0      & (\text{for} \ V_\mathrm{step}\geq gn_\mathrm{b})
    \end{cases}.
    \label{eq:TF desity profile}
\end{equation}
The numerical plots are well approximated by $\sqrt{n_\mathrm{TF}}$, except inside the potential at $V= gn_\mathrm{b}$ and in the vicinity of the interface.

According to Ref.~\cite{Takahashi2009a}, Eq.~\eqref{eq:time-independent GPE} has analytical solutions for all values of $V$ in an infinite-size system under the Neumann boundary conditions.
As shown in Fig.~\ref{fig:FIG1}(c), $\phi$ for $V\neq gn_\mathrm{b}$ exhibits a exponentially decaying profile with a decay length $\sim\xi(V)$ inside the step potential.
In contrast, the analytical solution for $V= gn_\mathrm{b}$ is $\phi(x)=\sqrt{n_\mathrm{b}}(x/\xi_\mathrm{b}+\sqrt{2})^{-1}$ inside the step potential, which indicates a power-law behavior $\phi\propto x^{-1}$ for $x\gg\xi_\mathrm{b}$.

We note that a potential height $V= gn_\mathrm{b}$ can be regarded as the critical point of the local condensation phase transition.
At $V= gn_\mathrm{b}$, the GP energy density inside the step potential (\textit{i.e.}, the integrand of Eq.~\eqref{eq:energy functional}) becomes so-called a Mexican hat potential.
Accordingly, $\phi$ indicates a power-law behavior for $x\gg\xi_\mathrm{b}$, and the healing length $\xi(V)$ diverges.
These behaviors are typically observed in critical phenomena~\cite{Stanley1987a,Landau2013b}.
In this paper, we refer to $V= gn_\mathrm{b}$ as the critical height.

\section{Lowest-energy excitations}\label{sec:lowest-energy excitations}
According to the Landau criterion of superfluidity, the critical velocity is determined by the dispersion relation of the lowest-energy excitation.
Before discussing the critical velocity in Sec.~\ref{sec:critical velocity}, we here present numerical results for the dispersion relations and wave functions of the excitations.

\begin{figure}[t]
\centering
\includegraphics[width=1.0\columnwidth]{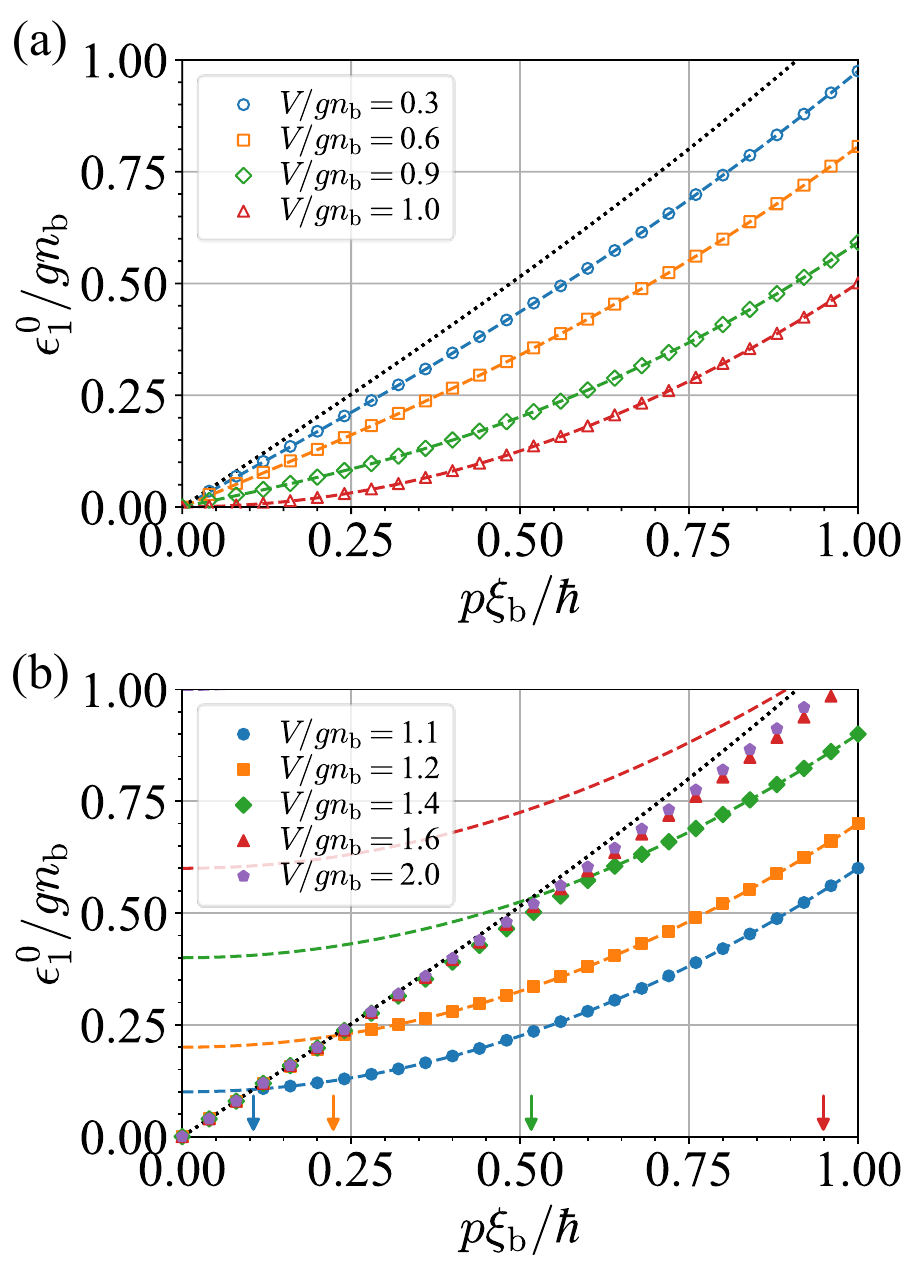}
\caption{
Dispersion relations $\epsilon_{1}^0(p)$ of the lowest-energy excitations for several values of (a) $V\leq gn_\mathrm{b}$ and (b) $V> gn_\mathrm{b}$.
Black dotted curves represent the Bogoliubov spectrum $\varepsilon_\mathrm{B}=\sqrt{\frac{p^2}{2m}(\frac{p^2}{2m}+2gn_\mathrm{b})}$ in the bulk.
Colored dashed curves represent the local Bogoliubov spectrum $\mathcal{E}_\mathrm{LB}$ [Eq.~\eqref{eq:LB spectrum}] in (a) and the gapful spectrum $\mathcal{E}_\mathrm{gap}$ [Eq.~\eqref{eq:gap spectrum}] in (b).
Colored arrows in (b) schematically indicate the values of $p_\mathrm{cross}$ [Eq.~\eqref{eq:pcross}], where $\varepsilon_\mathrm{B}$ and $\mathcal{E}_\mathrm{gap}$ cross.
}
\label{fig:FIG2}
\end{figure}

\subsection{Bogoliubov theory and Landau criterion}\label{sec:Bogoliubov theory}
In order to describe the elementary excitations of the stationary states, we consider a perturbed wave function $\Psi=\phi e^{im\bm{v}\cdot\bm{r}/\hbar}+\delta\Psi$.
The fluctuation is expressed as the collective excitation $\delta\Psi=\left[u_n(x)e^{i(\bm{p}\cdot\bm{r}-\epsilon_n t)/\hbar}-v_n^*(x)e^{-i(\bm{p}\cdot\bm{r}-\epsilon_n^* t)/\hbar}\right]e^{im\bm{v}\cdot\bm{r}/\hbar}$ with $\bm{p}=p\hat{\bm{r}}_\perp$.
By linearizing the equation of motion of the Lagrangian $\mathcal{L}$ with respect to $\delta\Psi$, we obtain the Bogoliubov-de Gennes (BdG) equation~\cite{Bogoliubov1947a}
\begin{equation}
    (\epsilon_n+vp)\vec{w}_n=\hat{H}\vec{w}_n
    \equiv
    \begin{bmatrix}
        h       & -g\phi^2 \\
        g\phi^2 & -h \\
    \end{bmatrix}
    \vec{w}_n
    \label{eq:BdGE}
\end{equation}
with $h=-\frac{\hbar^2}{2m}\frac{d^2}{dx^2}+\frac{p^2}{2m}+V_\mathrm{step}- gn_\mathrm{b}+2g\phi^2$ and $\vec{w}_n(x)=[u_n,v_n]^T$.
Here, $n-1$ denotes the number of nodes of the wave functions along the $x$-direction.
When the eigenvalue becomes real, the Bogoliubov coefficients are normalized as $\mathcal{N}_{nn}=1$ with $\mathcal{N}_{nn'}=\int d^3x \ \vec{w}_n^\dagger\hat{\sigma}_z\vec{w}_{n'}$ and $\hat{\sigma}_z=\mathrm{diag}(1,-1)$.
Then, we always have another eigensolution $(-\epsilon_n^0,v_n^*,u_n^*)$ with the negative norm $\int d^3x \ (|v_n|^2-|u_n|^2)=-\mathcal{N}_{nn}=-1$.
The excitation energies defined as $\epsilon_n^0\mathcal{N}_{nn}$ are the same for these two solutions, so they are physically identical.
Therefore, the solution with the negative norm is ignored in our analysis.
Furthermore, since $\hat{H}$ is a real matrix, the eigenvector $\vec{w}_n$ can be taken to be real by an appropriate choice of the global phase.

According to Eq.~\eqref{eq:BdGE}, the Doppler shift for the dispersion relation is given by $\epsilon_n(p)=\epsilon_n^0(p)-vp$.
Here, $\epsilon_n^0$ is the dispersion for $\bm{v}=0$.
If $\epsilon_n<0$ for at least one mode, the stationary state becomes thermodynamically unstable.
In that case, excitations with the negative energy are spontaneously emitted to reduce the thermodynamic energy of the system.
When $p\neq0$, the total momentum of the condensate decreases, thereby leading to breakdown of superfluidity.
According to the Landau criterion of superfluidity~\cite{Landau1941a}, the flow becomes dissipative when $v$ exceeds a critical velocity 
\begin{equation}
    v_\mathrm{c}
    =\min_p\left[\frac{\epsilon_1^0(p)}{p}\right]
    =\frac{\epsilon_1^0(p_\mathrm{c})}{p_\mathrm{c}}
    \label{eq:critical velocity}
\end{equation}
with a critical momentum $p_\mathrm{c}$.
Hence, $v_\mathrm{c}$ is determined solely by the dispersion relation $\epsilon_1^0(p)$ of the lowest-energy excitation.
In the following sections, we numerically and theoretically investigate $\epsilon_1^0$ and $\vec{w}_1=[u_1,v_1]^T$.

\begin{figure}[t]
\centering
\includegraphics[width=1.0\columnwidth]{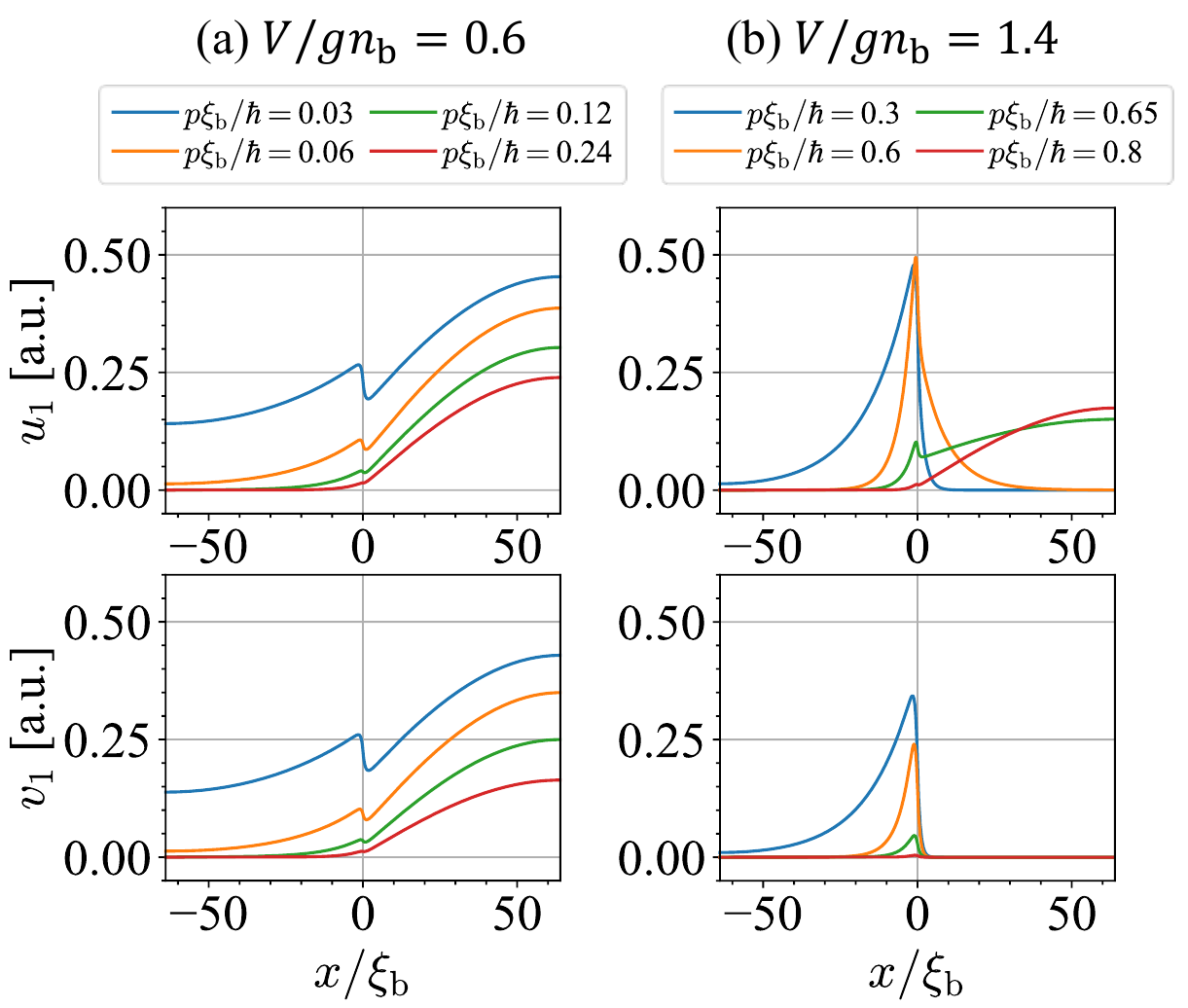}
\caption{
Typical profiles of the wave functions $u_1$ (top) and $v_1$ (bottom) of the low-energy excitations for (a) $V=0.6 gn_\mathrm{b}$ and (b) $V=1.4 gn_\mathrm{b}$.
At $p=0$, both $u_1$ and $v_1$ of the NG mode are proportional to the stationary state $\phi$.
}
\label{fig:FIG3}
\end{figure}

\subsection{Numerical results for lowest-energy excitations }\label{sec:numerical results for lowest-energy excitations}
In Fig.~\ref{fig:FIG2}, we present the dispersion relations of the lowest-energy excitations for several values of $V$.
They are obtained by numerically diagonalizing Eq.~\eqref{eq:BdGE}, as detailed in Appendix~\ref{app:numerical methods}.
At $p=0$, a gapless mode called the Nambu-Goldstone (NG) mode~\cite{Nambu1961a,Goldstone1961a,Goldstone1962a} associated with the spontaneous breaking of the U(1) symmetry of $\phi$ exists.
In the absence of the step potential ($V=0$), the system becomes uniform and thus the dispersion coincides with the Bogoliubov spectrum $\varepsilon_\mathrm{B}(p)=\sqrt{\frac{p^2}{2m}(\frac{p^2}{2m}+2gn_\mathrm{b})}$~\cite{Bogoliubov1947a}.
As discussed in Sec.~\ref{sec:critical velocity}, $\epsilon_1^0$ for $V\leq gn_\mathrm{b}$ agrees well with the local Bogoliubov spectrum inside the potential [Fig.~\ref{fig:FIG2}(a)].
As $V$ exceeds $ gn_\mathrm{b}$, however, the plot of $\epsilon_1^0$ has a kink, above which the dispersion becomes quadratic [Fig.~\ref{fig:FIG2}(b)].
As $V$ increases further, the kink becomes less pronounced and the dispersion gradually approaches $\varepsilon_\mathrm{B}$.

Typical profiles of the wave functions $u_1$ and $v_1$ for $V< gn_\mathrm{b}$ and $V> gn_\mathrm{b}$ are plotted in Fig.~\ref{fig:FIG3}.
For the NG mode, $u_1=v_1=i\frac{\Delta\Theta}{2}\phi$ is satisfied under an infinitesimal global phase rotation $\phi\to e^{i\Delta\Theta}\phi$.
At $p\neq0$, the profiles of $u_1$ and $v_1$ for $V< gn_\mathrm{b}$ are similar and become sinusoidal (decaying) forms inside (outside) the potential [Fig.~\ref{fig:FIG3}(a)].
Near the interface, $|\frac{du_1}{dx}|$ and $|\frac{dv_1}{dx}|$ become larger than in other regions.
For $V> gn_\mathrm{b}$, however, the amplitude of $v_1$ vanishes inside the potential [Fig.~\ref{fig:FIG3}(b)]. 
Furthermore, we find that some wave functions are localized near the interface (see $p\xi_\mathrm{b}/\hbar=0.3$ and $0.6$), which are never observed for $V< gn_\mathrm{b}$.

\begin{figure*}[t]
\centering
\includegraphics[width=1.0\textwidth]{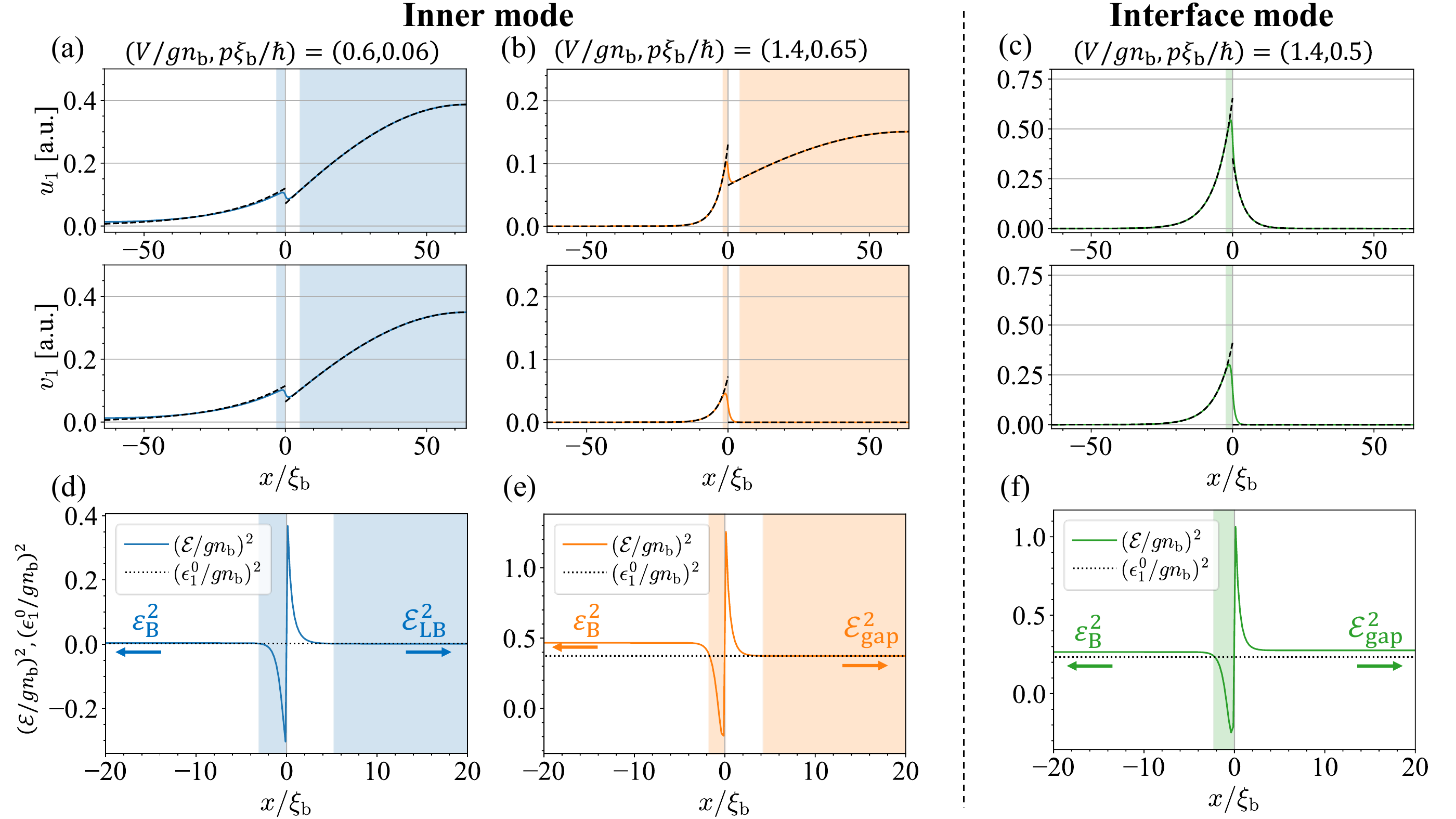}
\caption{
Comparison between the inner mode and interface mode within the semi-classical theory. 
The top and middle panels show typical profiles of the wave functions $u_1$ and $v_1$, respectively, for (a) $(V/ gn_\mathrm{b},p\xi_\mathrm{b}/\hbar)=(0.6,0.06)$, (b) $(V/ gn_\mathrm{b},p\xi_\mathrm{b}/\hbar)=(1.4,0.65)$, and (c) $(V/ gn_\mathrm{b},p\xi_\mathrm{b}/\hbar)=(1.4,0.5)$.
Colored solid curves represent the numerical plots, while black dashed curves represent Eqs.~\eqref{eq:cos WKB solution} and \eqref{eq:exp WKB solution}.
The bottom panels (d), (e), and (f) show the corresponding profiles of the square of the effective energy barrier $\mathcal{E}^2$ [Eq.~\eqref{eq:effective energy barrier}].
Black dotted lines represent $(\epsilon_1^0)^2$.
Rightward arrows indicate $\mathcal{E}^2\to\mathcal{E}_\mathrm{LB}^2$ [Eq.~\eqref{eq:LB spectrum}] and $\mathcal{E}^2\to\mathcal{E}_\mathrm{gap}^2$ [Eq.~\eqref{eq:gap spectrum}] far from the interface, whereas leftward arrows indicate $\mathcal{E}^2\to\varepsilon_\mathrm{B}^2$ with the Bogoliubov spectrum $\varepsilon_\mathrm{B}$.
For clarity, the regions where $\bar{P}^2(x)>0$ or $(\epsilon_1^0)^2>\mathcal{E}^2$ are shaded in light colors in all panels.
}
\label{fig:FIG4}
\end{figure*}

\section{Semi-classical analysis}\label{sec:semi-classical analysis}
Although the numerical results presented in Sec.~\ref{sec:numerical results for lowest-energy excitations} show the nontrivial behavior of the lowest-energy excitations at $p\neq0$, their physical interpretation cannot be fully obtained from these results alone.
To gain deeper insight, we introduce the semi-classical theory to extend our investigation of the excitations.

As shown in Sec.~\ref{sec:profiles of the stationary states}, the spatial variation of $\phi$ is sufficiently smooth except in the vicinity of the interface.
In this case, the semi-classical theory for the BdG equation provides a good approximation as shown below.
The Wentzel-Kramers-Brillouin (WKB) approximation has been successfully applied to bosonic quasiparticles in various contexts, such as the analogue Hawking radiation in BECs~\cite{Coutant2010a,Finazzi2010a,Wang2017a}, and dynamic instability of a doubly quantized vortex~\cite{Takeuchi2018a} and flat domain walls in multicomponent BECs~\cite{Takeuchi2013a,Takeuchi2022a}.
Based on the semi-classical analysis for the Schr\"{o}dinger equation~\cite{Landau2013a}, we here develop a general formulation applicable to our problem.

We start from the WKB ansatz $\vec{w}_n(x)=e^{iS/\hbar}\vec{\mathcal{W}}_n$ with coefficients $\vec{\mathcal{W}}_n=[\mathcal{U}_n,\mathcal{V}_n]^T$.
By expanding $S(x)$ in powers of $\hbar$ as $S=S_0+\frac{\hbar}{i}S_1$, the BdG equation $\epsilon_n^0\vec{w}_n=\hat{H}\vec{w}_n$ reduces to 
\begin{equation}
    \epsilon_n^0\vec{\mathcal{W}}_n=
    \begin{bmatrix}
        h_0     & -g\phi^2 \\
        g\phi^2 & -h_0 \\
    \end{bmatrix}
    \vec{\mathcal{W}}_n
    +\frac{\hbar}{i}
    \begin{bmatrix}
        D & 0 \\
        0 & -D \\
    \end{bmatrix}
    \vec{\mathcal{W}}_n
    \label{eq:WKB BdGE}
\end{equation}
with $P(x)=\frac{dS_0}{dx}$, $h_0(x)=\frac{P^2+p^2}{2m}+V_\mathrm{step}- gn_\mathrm{b}+2g\phi^2$, and $D(x)=\frac{P
}{m}\frac{dS_1}{dx}+\frac{1}{2m}\frac{dP}{dx}$.
The spatial derivative is replaced as $-\frac{\hbar^2}{2m}\frac{d^2}{dx^2}\vec{w}_n\to e^{iS/\hbar}\frac{1}{2m}(S'^2+\frac{\hbar}{i}S'')\vec{\mathcal{W}}_n$.
As in the case of the Schr\"{o}dinger equation~\cite{Landau2013a}, the approximation used in this expression is valid under $\hbar|S''/S'^2|\ll1$, or equivalently $|\frac{d(\hbar/P)}{dx}|\ll1$ in the classical limit $\hbar\to0$.
Therefore, the WKB approximation works well except in the vicinity of the interface, where $V_\mathrm{step}$ and $\phi$ vary rapidly.

In the zeroth-order approximation, we ignore the second term on the right side of Eq.~\eqref{eq:WKB BdGE} and obtain $(\epsilon_n^0-h_0)(\epsilon_n^0+h_0)+(g\phi^2)^2=0$
Solving this equation with respect to $P^2$ yields two solutions $\pm\bar{P}^2$, where 
\begin{equation}
    \bar{P}^2(x)
    =2m\frac{(\epsilon_n^0)^2-\mathcal{E}^2}{\sqrt{(\epsilon_n^0)^2+(g\phi^2)^2}+\sqrt{\mathcal{E}^2+(g\phi^2)^2}}
    \label{eq:classical momentum}
\end{equation}
with the square of a effective energy barrier for elementary excitations, 
\begin{align}
    &\mathcal{E}^2(p,V_\mathrm{step}(x),\phi^2(x)) \notag \\
    &\equiv\left(
      \frac{p^2}{2m}+V_\mathrm{step}- gn_\mathrm{b}+2g\phi^2
     \right)^2
     -(g\phi^2)^2.
    \label{eq:effective energy barrier}
\end{align}
The plus and minus signs of $\pm\bar{P}^2$ correspond to eigenmodes with positive and negative norms, respectively; therefore, we choose the plus sign.
The first-order correction satisfies $D=0$ and reduces to $\frac{dS_1}{dx}=-\frac{1}{2}\frac{d\ln P}{dx}=0$.
Finally, we obtain the general solution as a linear combination with respect to $\pm\bar{P}$,
\begin{equation}
    \vec{w}_n(x)
    =\left(
     \frac{C_+}{\sqrt{|\bar{P}|}}e^{\frac{i}{\hbar}\int\bar{P}\,dx}+
     \frac{C_-}{\sqrt{|\bar{P}|}}e^{-\frac{i}{\hbar}\int\bar{P}\,dx}
     \right)\vec{\mathcal{W}}_n
     \label{eq:WKB solution}
\end{equation}
with complex constants $C_+$ and $C_-$.

To analyze the behavior of the excitations far from the interface, we combine the semi-classical theory with the TF approximation [Eq.~\eqref{eq:TF desity profile}].
According to Eq.~\eqref{eq:effective energy barrier}, we obtain $\mathcal{E}^2(p,V\leq gn_\mathrm{b},n_\mathrm{TF})=\mathcal{E}^2_\mathrm{LB}$ and $\mathcal{E}^2(p,V\geq gn_\mathrm{b},n_\mathrm{TF})=\mathcal{E}^2_\mathrm{gap}$ inside the potential.
Here, we define the local Bogoliubov spectrum 
\begin{equation}
    \mathcal{E}_\mathrm{LB}(p)
    \equiv\sqrt{\frac{p^2}{2m}
    \left\{
    \frac{p^2}{2m}+2(gn_\mathrm{b}-V)
    \right\}}
    \label{eq:LB spectrum}
\end{equation}
and the gapful spectrum
\begin{equation}
    \mathcal{E}_\mathrm{gap}(p)
    \equiv\frac{p^2}{2m}+V- gn_\mathrm{b}.
    \label{eq:gap spectrum}
\end{equation}
Outside the potential, we obtain $\mathcal{E}^2(p,0,n_\mathrm{b})=\varepsilon_\mathrm{B}^2$ with the Bogoliubov spectrum $\varepsilon_\mathrm{B}$ in the bulk.
In the TF approximation, Eq.~\eqref{eq:classical momentum} becomes constant as $\bar{P}(x)\to\bar{P}_\mathrm{in}$ inside the potential and $\bar{P}(x)\to\bar{P}_\mathrm{out}$ outside the potential.
If $\bar{P}_\mathrm{in,out}^2>0$, it is straightforward to prove that Eq.~\eqref{eq:WKB solution} reduces to a oscillatory solution
\begin{equation}
    \vec{w}_n(x)
    =\vec{\mathcal{W}}_n
     \cos\left(
      \frac{\bar{P}_\mathrm{in,out}x}{\hbar}+\theta_n
     \right).
    \label{eq:cos WKB solution}
\end{equation}
Here, the coefficients $|\bar{P}_\mathrm{in,out}|^{-1/2}$ are included in $\vec{\mathcal{W}}_n$ and $\theta_n$.
In the same manner, if $\bar{P}_\mathrm{in,out}^2<0$, we have a exponential solution 
\begin{equation}
    \vec{w}_n(x)
    =\vec{\mathcal{W}}_n 
     \exp\left(
      {\mp\frac{| \bar{P}_\mathrm{in,out}|x}{\hbar}}
     \right)
    \label{eq:exp WKB solution},
\end{equation}
where the minus and plus signs correspond to the indices ``in'' and ``out'', respectively.
Furthermore, the ratio of $\mathcal{U}_n$ to $\mathcal{V}_n$ is given by 
\begin{equation}
    \frac{\mathcal{V}_n}{\mathcal{U}_n}
    =\frac{gn_\mathrm{TF}}{\sqrt{(\epsilon_n^0)^2+(gn_\mathrm{TF})^2}+\epsilon_n^0},
    \label{eq:ratio}
\end{equation}
which implies $\mathcal{V}_n=0$ when $n_\mathrm{TF}=0$, consistent with the numerical results shown in Fig.~\ref{fig:FIG3}(b). 
In the following analysis, we focus on the above discussion for $n=1$~\footnote{
Our formulation of the semi-classical theory works well for $n\geq2$, although figures analogous to Fig.~\ref{fig:FIG4} are not shown in this paper.
}.
As explained below, the excitations at $p\neq0$ can be classified into inner modes and interface modes.

The inner modes are excited for $V\leq gn_\mathrm{b}$ at $p\neq0$, and for $ gn_\mathrm{b}<V<2 gn_\mathrm{b}$ at $p\gtrsim p_\mathrm{cross}$.
Here, $\varepsilon_\mathrm{B}$ and $\mathcal{E}_\mathrm{gap}$ cross at $p=p_\mathrm{cross}$ with 
\begin{equation}
    p_\mathrm{cross}
    =\hbar\xi_\mathrm{b}^{-1}
    \frac{V/ gn_\mathrm{b}-1}{\sqrt{2-V/ gn_\mathrm{b}}} \quad 
    (\text{for} \  gn_\mathrm{b}\leq V<2 gn_\mathrm{b}).
    \label{eq:pcross}
\end{equation}
Figures~\ref{fig:FIG4}(a) and (b) show typical profiles of $u_1$ and $v_1$ for the inner modes.
The term ``inner'' refers to a mode that extends inside the step potential and not inside the bulk of the superfluid.
Except in the vicinity of the interface, the numerical plots are well fitted by the semi-classical descriptions with Eq.~\eqref{eq:cos WKB solution} inside the potential and Eq.~\eqref{eq:exp WKB solution} outside the potential~\footnote{
Under the Neumann boundary condition at $x=L_x/2$, we have $\theta_1=-\bar{P}_\mathrm{in}L_x/(2\hbar)$ and $\mathcal{U}_1=u_1(L_x/2)$ in Eq.~\eqref{eq:cos WKB solution}.
In contrast, Eq.~\eqref{eq:exp WKB solution} cannot satisfy the boundary conditions exactly.
Then, we use $u_1(-L_x/4)$ to determine $\mathcal{U}_1$ as $\mathcal{U}_1=u_1(-L_x/4)e^{|\bar{P}_\mathrm{out}|L_x/(4\hbar)}$.
$\mathcal{V}_1$ is determined by the ratio [Eq.~\eqref{eq:ratio}].
For the interface modes, the coefficient $\vec{\mathcal{U}}_1$ in Eq.~\eqref{eq:exp WKB solution} is determined in the same manner as above.
}. 
As shown in Figs.~\ref{fig:FIG4}(d) and (e), $(\epsilon_1^0)^2$ is larger than $\mathcal{E}^2$ inside the potential, leading to a real momentum $\bar{P}$ in the $x$-direction.
In contrast, $(\epsilon_1^0)^2$ is smaller than $\mathcal{E}^2$ and $\bar{P}^2$ becomes negative outside the potential, where the excitations are classically forbidden.
Near the interface, however, $\bar{P}^2$ becomes positive (negative) outside (inside) the potential.
As a result, the excitation acquires a momentum corresponding to a wavelength of order $\xi_\mathrm{b}$, which allows the wave functions to vary rapidly and remain continuous across the interface.

The interface modes are excited for $ gn_\mathrm{b}<V<2 gn_\mathrm{b}$ at $p\lesssim p_\mathrm{cross}$, and for $V\geq2 gn_\mathrm{b}$ at $p\neq0$. 
Figure~\ref{fig:FIG4}(c) shows that $u_1$ and $v_1$ exhibit sharp peaks near the interface.
Far from the interface, $u_1$ decays exponentially because $\bar{P}^2$ becomes negative and the excitations are classically forbidden (see Fig.~\ref{fig:FIG4}(f)).
Furthermore, $\bar{P}^2$ becomes positive near the interface in $x<0$, where the bound state are locally formed.
We note that Bogoliubov modes at interfaces play a crucial role in other contexts, as for example vortex nucleation near the surface of large BECs, as shown in Ref.~\cite{Anglin2001a}.
If the decay length $\hbar|\bar{P}_\mathrm{out,in}|^{-1}$ becomes larger, the interface modes gradually change into the bulk modes with the Bogoliubov spectrum $\varepsilon_\mathrm{B}$.
We note that the interface modes for $ gn_\mathrm{b}<V<2 gn_\mathrm{b}$ are considered to be associated with an avoided crossing~\cite{Greiner2011a} between two branches $\varepsilon_\mathrm{B}$ and $\mathcal{E}_\mathrm{gap}$.
In Fig.~\ref{fig:FIG2}(b), we find that $\epsilon_1^0$ lies below $\varepsilon_\mathrm{B}$ and $\mathcal{E}_\mathrm{gap}$ owing to the avoided crossing around $p=p_\mathrm{cross}$. 
Therefore, the coupling between these branches lowers $\epsilon_1^0$, thereby lifting the degeneracy between them.

\begin{figure}[t]
\centering
\includegraphics[width=1.0\columnwidth]{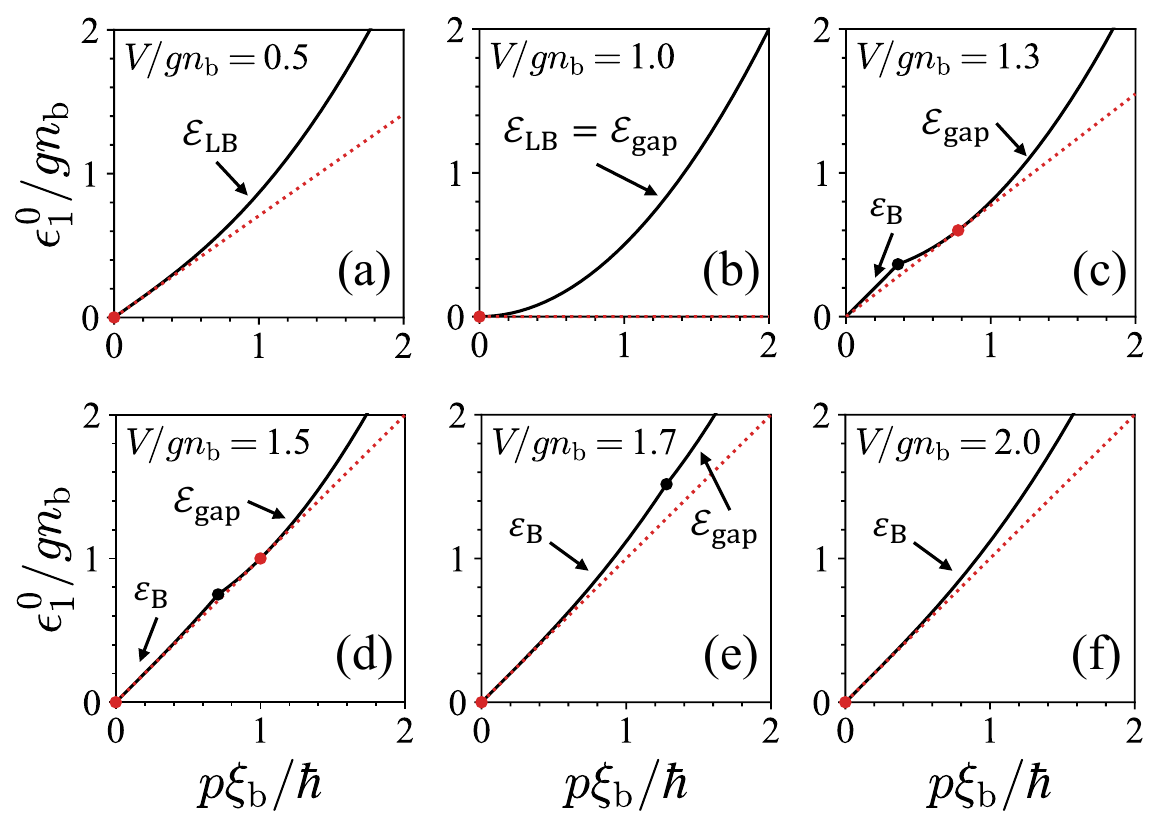}
\caption{
Theoretical dispersion relations in an infinite system for several values of $V$.
Red dotted lines $\epsilon_1^0=v_\mathrm{c}p$ are tangent to black solid curves at $(p_\mathrm{c},v_\mathrm{c}p_\mathrm{c})$ as indicated by red markers. 
Black markers in (c)-(e) represent the crossing points at which $\varepsilon_\mathrm{B}(p_\mathrm{cross})=\mathcal{E}_\mathrm{gap}(p_\mathrm{cross})$ with $p_\mathrm{cross}$ [Eq.~\eqref{eq:pcross}].
}
\label{fig:FIG5}
\end{figure}

\section{Critical velocity}\label{sec:critical velocity}
The semi-classical theory successfully accounts for the $V$-dependence of the critical velocity $v_\mathrm{c}$ in an infinite system.
According to Eq.~\eqref{eq:critical velocity}, $v_\mathrm{c}$ is determined by the dispersion relation $\epsilon_1^0(p)$.
For $V\leq gn_\mathrm{b}$, Fig.~\ref{fig:FIG2}(a) demonstrates that $\epsilon_1^0$ agrees well with the local Bogoliubov spectrum $\mathcal{E}_\mathrm{LB}$ [Eq.~\eqref{eq:LB spectrum}].
Therefore, the theoretical dispersion is considered to be $\epsilon_1^0=\mathcal{E}_\mathrm{LB}(p)$, as shown in Figs.~\ref{fig:FIG5}(a) and (b).
For $ gn_\mathrm{b}<V<2 gn_\mathrm{b}$, Fig.~\ref{fig:FIG2}(b) demonstrates that $\epsilon_1^0$ agrees well with the Bogoliubov spectrum $\varepsilon_\mathrm{B}$ for $p\lesssim p_\mathrm{cross}$ and with the gapful spectrum $\mathcal{E}_\mathrm{gap}$ [Eq.~\eqref{eq:gap spectrum}] for $p\gtrsim p_\mathrm{cross}$.
Neglecting the avoided crossing around $p=p_\mathrm{cross}$, the theoretical dispersion consists of two branches; 
$\epsilon_1^0=\varepsilon_\mathrm{B}(p)$ for $p\leq p_\mathrm{cross}$ and $\epsilon_1^0=\mathcal{E}_\mathrm{LB}(p)$ for $p\geq p_\mathrm{cross}$, as shown in Figs.~\ref{fig:FIG5}(c)-(e).
For $V\geq2 gn_\mathrm{b}$, the numerical plots of $\epsilon_1^0$ are well fitted by $\varepsilon_\mathrm{B}$ for low momentum.
We therefore assume $\epsilon_1^0=\varepsilon_\mathrm{B}(p)$, as shown in Fig.~\ref{fig:FIG5}(f).
As a result, we obtain the following expression for the critical velocity;
\begin{equation}
    v_\mathrm{c}=
    \begin{cases}
        c_\mathrm{s}\sqrt{1-V/ gn_\mathrm{b}}    & (\text{for} \ 0\leq V\leq gn_\mathrm{b}) \\
        c_\mathrm{s}\sqrt{2(V/ gn_\mathrm{b}-1)} & (\text{for} \  gn_\mathrm{b}\leq V\leq3gn_\mathrm{b}/2) \\
        c_\mathrm{s}                    & (\text{for} \ V\geq3gn_\mathrm{b}/2)
    \end{cases},
    \label{eq:theoretical critical velocity}
\end{equation}
where $c_\mathrm{s}=\sqrt{gn_\mathrm{b}/m}$ is the speed of sound in the bulk.
Here, the critical momentum is $p_\mathrm{c}=\hbar\xi_\mathrm{b}^{-1}\sqrt{2(V/ gn_\mathrm{b}-1)}$ for $ gn_\mathrm{b}\leq V\leq3gn_\mathrm{b}/2$ and $p_\mathrm{c}=0$ otherwise.

\begin{figure}[t]
\centering
\includegraphics[width=1.0\columnwidth]{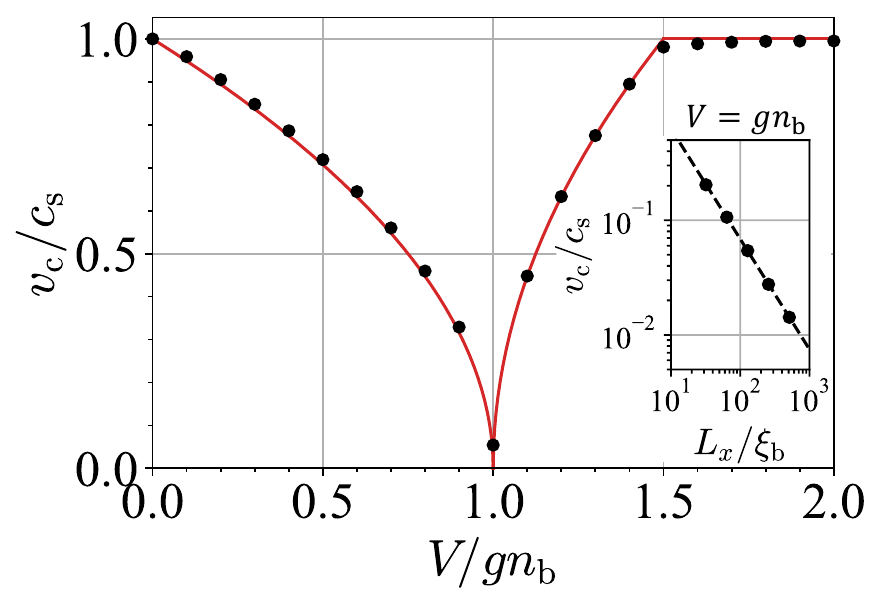}
\caption{
Critical velocity $v_\mathrm{c}$ versus the potential height $V$.
Red solid curves represent $v_\mathrm{c}$ [Eq.~\eqref{eq:theoretical critical velocity}] obtained theoretically in an infinite system.
Black markers are obtained numerically in a finite system with $L_x=128\xi_\mathrm{b}$.
The inset shows the $L_x$-dependence of $v_\mathrm{c}$ at the critical height $V= gn_\mathrm{b}$. 
The black dashed line indicates a power-law function $v_\mathrm{c}/c_\mathrm{s}=\beta(L_x/\xi_\mathrm{b})^\gamma$ with fitting parameters $\beta\approx5.779$ and $\gamma\approx-0.963$.
}
\label{fig:FIG6}
\end{figure}

Figure~\ref{fig:FIG6} clearly shows that the numerical results for $v_\mathrm{c}$ are in good agreement with Eq.~\eqref{eq:theoretical critical velocity}.
We find that $v_\mathrm{c}$ is sharply minimized at the critical height $V= gn_\mathrm{b}$.
In the following, we illustrate the mechanism underlying this minimization graphically.
According to Eq.~\eqref{eq:critical velocity}, the critical velocity corresponds to the slope of the line $\epsilon_1^0=v_\mathrm{c}p$ that is tangent to the curve of the dispersion at $(p_\mathrm{c},v_\mathrm{c}p_\mathrm{c})$.
As shown in Figs.~\ref{fig:FIG5}(a) and (b) for $0\leq V\leq gn_\mathrm{b}$, the local Bogoliubov spectrum $\epsilon_1^0=\mathcal{E}_\mathrm{LB}(p)$ provides $v_\mathrm{c}=\sqrt{gn_\mathrm{TF}/m}$ with $p_\mathrm{c}=0$.
Here, $\sqrt{gn_\mathrm{TF}/m}$ is the local speed of sound with the local density inside the potential.
As $V$ increases toward $ gn_\mathrm{b}$, $v_\mathrm{c}=c_\mathrm{s}\sqrt{1-V/ gn_\mathrm{b}}$ decreases and eventually becomes zero.
For $ gn_\mathrm{b}<V<3gn_\mathrm{b}/2$, Fig.~\ref{fig:FIG5}  (c) demonstrates that $\epsilon_1^0=v_\mathrm{c}p$ is tangent to $\epsilon_1^0=\mathcal{E}_\mathrm{gap}(p)$ at $p=p_\mathrm{c}>p_\mathrm{cross}>0$.
The gapful spectrum $\mathcal{E}_\mathrm{gap}$ has a energy gap $V- gn_\mathrm{b}$ originating from the effective energy barrier inside the potential, given by Eq.~\eqref{eq:effective energy barrier}.
We reveal that this energy gap results in the recovery of the critical velocity from zero as $v_\mathrm{c}=c_\mathrm{s}\sqrt{2(V/ gn_\mathrm{b}-1)}$.
For $V\geq3gn_\mathrm{b}/2$, Figs.~\ref{fig:FIG5}(d)-(f) show that $\epsilon_1^0=c_\mathrm{s}p$ is tangent to the Bogoliubov spectrum $\epsilon_1^0=\varepsilon_\mathrm{B}(p)$ at $p=p_\mathrm{c}=0$, consistent with $v_\mathrm{c}=c_\mathrm{s}$ in the hard-wall limit $V\to\infty$~\cite{Kovrizhin2001a}.
We note that the numerical results around $V=3gn_\mathrm{b}/2$ are slightly smaller than $v_\mathrm{c}=c_\mathrm{s}$ due to the avoided crossing
~\footnote{
Numerical results show that the interface modes are excited at $p=p_\mathrm{c}$ for $3 gn_\mathrm{b}/2<V<2 gn_\mathrm{b}$ due to the avoided crossing
}.

Finally, we examine finite-size effects on the critical velocity at the critical height $V= gn_\mathrm{b}$.
As shown in the inset of Fig.~\ref{fig:FIG6}, we find a power-law scaling $v_\mathrm{c}/c_\mathrm{s}=\beta(L_x/\xi_\mathrm{b})^\gamma $ with fitting parameters $\beta\approx5.779$ and $\gamma\approx-0.963$.
Here, $\beta$ and $\gamma$ are determined by performing the least squares method to the log-log plot of the data.
In the infinite-size limit $L_x\to\infty$, $v_\mathrm{c}/c_\mathrm{s}\to0$ is in reasonable agreement with the theoretical prediction given by Eq.~\eqref{eq:theoretical critical velocity}.
This power-law scaling of $v_\mathrm{c}$ clearly reflects the criticality at $V= gn_\mathrm{b}$.
Furthermore, we have checked that scaling only the obstacle size at a fixed system size results also in a power law scaling of $v_\mathrm{c}$, as can be expected.

\section{Summary and Discussion}\label{sec:summary}
We theoretically investigated the critical velocity $v_\mathrm{c}$ of a BEC flowing along a step potential.
By analyzing the $V$-dependence of the stationary states $\phi$, we demonstrated that the potential height $V= gn_\mathrm{b}$ corresponds to the critical point of the local condensation phase transition inside the step potential.
At the critical height, $\phi$ exhibits a power-law decay inside the potential [Fig.~\ref{fig:FIG1}(c)]. 
Our semi-classical analysis explains the numerical results for the dispersion relations [Fig.~\ref{fig:FIG2}] and wave functions of the lowest-energy excitations [Fig.~\ref{fig:FIG4}] very well.
We theoretically showed that $v_\mathrm{c}$ is sharply minimized and becomes zero at $V= gn_\mathrm{b}$, consistent with the power-law scaling $v_\mathrm{c}\propto L_x^{-0.963}$ with the system size $L_x$ [Fig.~\ref{fig:FIG6}].
For $V< gn_\mathrm{b}$, $v_\mathrm{c}$ equals the local speed of sound inside the potential and approaches zero as the local density decreases.
For $ gn_\mathrm{b}<V<3gn_\mathrm{b}/2$, the energy gap $V- gn_\mathrm{b}$ of the gapful spectrum leads to the recovery of $v_\mathrm{c}$ from zero.
When $V$ exceeds $3gn_\mathrm{b}/2$, $v_\mathrm{c}$ reaches and converges to the speed of sound in the bulk.

The superfluid critical velocity has traditionally been understood using the dispersion relation for elementary excitations, as expressed in Eq.~\eqref{eq:critical velocity}, called the Landau criterion. 
To the best of our knowledge, this study is the first to provide such a microscopic explanation, using the dispersion relation of Bogoliubov excitations, for the anomalous minimization of the critical velocity observed in Ref.~\cite{Kwon2015a}. 
Such a microscopic explanation not only deepens our understanding of superfluidity at a fundamental level but is also expected to be applicable to general cases; 
for example, our theory could explain qualitatively the critical velocity for macroscopic obstacles of arbitrary shapes with smooth surfaces and no sharp corners.

The experiment by Kwon \textit{et al.}~\cite{Kwon2015a} provides support for our results.
By using a two-dimensional Gaussian potential $V(x,y)=V_0\exp[-2(x^2+y^2)/\sigma^2]$, they observed a sharp minimum of $v_\mathrm{c}$ at $V_0\approx\mu$ and a power-law scaling as $v_\mathrm{c}\sim\sigma^{-0.78}$~\cite{Kwon2015a,Kwak2023a}.
The differences between the step and Gaussian potentials are reflected in the following aspects:
the convergence of $v_\mathrm{c}$ toward a constant value of $\sim0.3c_\mathrm{s}$ for $V_0\gg\mu$, a slight deviation of the critical height from $\mu$, and values of the critical exponent of the power-law scaling.
To interpolate between the step and Gaussian potentials, a natural extension of this work is to consider a rectangular wall potential with a finite width.
Furthermore, our semi-classical analysis can be straightforwardly generalized to two dimensions.
In this way, the present approach can be extended to provide a useful framework for analyzing more realistic experimental configurations based on recent optical techniques, including  box potentials~\cite{Gaunt2013a,Navon2021a,Kwon2021a,Hernandez2024a} and DMDs~\cite{Gauthier2016a,Navon2021a}.


\section*{Acknowledgements}
This work was supported by JST, PRESTO Grant No. JPMJPR23O5, Japan, JSPS KAKENHI Grant Nos. JP20H01842, JP25K07187, and JP26K00640.

\section*{Data availability}
The data that support the findings of this article are not publicly available upon publication because it is not technically feasible and/or the cost of preparing, depositing, and hosting the data would be prohibitive within the scope of this research project.
The data are available from the authors upon reasonable request.



\appendix
\section{Numerical methods}\label{app:numerical methods}
We explain the numerical methods employed in this study. 
All simulations are performed by rescaling length, time, and wave function with $\xi_\mathrm{b}$, $\hbar/ gn_\mathrm{b}$, and $\sqrt{n_\mathrm{b}}$, respectively.

Numerical solutions of the stationary state $\phi$ are obtained by minimizing the GP energy functional [Eq.~\eqref{eq:energy functional}].
We use the steepest descent method to solve the imaginary time propagation $\frac{\partial\Psi}{\partial\tau}=-\frac{\delta \mathcal{F}}{\delta\Psi^*}$ with the imaginary time $\tau=it$.
The time evolution is written as $\Psi(n_\tau+1)=\Psi(n_\tau)-\Delta\tau\frac{\delta \mathcal{F}}{\delta\Psi^*}(n_\tau)$ with the discretized time $\tau=n_\tau\Delta\tau \ (n_\tau=0,1,2,\cdots)$.
The iteration of the evolutions is continued until $|\Psi(n_\tau+1)-\Psi(n_\tau)|$ converges within the double-precision accuracy, using the Intel Fortran Compiler.
After the finial step of the iteration, we obtain the stationary solution $\phi=\Psi(n_\tau+1)$.
At $x=\pm L_x/2$, the Neumann boundary conditions $\frac{d\Psi}{dx}|_{x=\pm L_x/2}=0$ are imposed.
Our simulations are performed on a one-dimensional spatial grid $x$, discretized as $x\to x_i=-L_x/2+(i-1/2)\Delta x \ (i=0,1,\cdots,N_x+1)$ with $L_x=N_x\Delta x$, $N_x=512$, and $\Delta x=0.25\xi_\mathrm{b}$.
The spatial derivative of $\Psi$ is computed using the central difference approximation $\frac{d^2\Psi}{dx^2}\to\frac{\Psi(x_{i-1})-2\Psi(x_i)+\Psi(x_{i+1})}{(\Delta x)^2}$.
The potential height $V$ is varied in the range $0\leq V\leq2 gn_\mathrm{b}$.

We obtain the dispersion relation $\epsilon_n^0(p)$ of the elementary excitations by using the numerical results for $\phi$ and numerically diagonalizing the discretized matrix of $\hat{H}$ with respect to the eigenvector $\vec{w}_n(x)=[u_n,v_n]^T$.
The diagonalization is performed by using the Intel Fortran Compiler with the Linear Algebra PACKage (LAPACK).
The discretization of the spatial coordinate and the boundary conditions on $u_n$ and $v_n$ are the same as above. 
The momentum of the elementary excitations is varied as $p=p_j=j\Delta p \ (j=0,1,2,\cdots)$ with $\Delta p=0.01\hbar\xi_\mathrm{b}^{-1}$.


\bibliography{reference}

\end{document}